 \documentclass[twocolumn,prl,aps,psfig,showpacs,preprintnumbers,superscriptaddress]{revtex4}
\usepackage{graphicx}
\usepackage{epstopdf}
\usepackage{float}
\usepackage{color}
\usepackage{amsmath}

\begin{document}

\title{NMR Determination of an Incommensurate Helical Antiferromagnetic Structure in EuCo$_2$As$_2$ }
\author{Q.-P. Ding}
\affiliation{Ames Laboratory, U.S. DOE, and Department of Physics and Astronomy, Iowa State University, Ames, Iowa 50011, USA}
\author{N. Higa}
\affiliation{Ames Laboratory, U.S. DOE, and Department of Physics and Astronomy, Iowa State University, Ames, Iowa 50011, USA}
\affiliation{Department of Physics and Earth Sciences, Faculty of Science, University of the Ryukyus, Okinawa 903-0213, Japan}
\author{N.~S.~Sangeetha}
\affiliation{Ames Laboratory, U.S. DOE, and Department of Physics and Astronomy, Iowa State University, Ames, Iowa 50011, USA}
\author{D. C. Johnston}
\affiliation{Ames Laboratory, U.S. DOE, and Department of Physics and Astronomy, Iowa State University, Ames, Iowa 50011, USA}
\author{Y. Furukawa}
\affiliation{Ames Laboratory, U.S. DOE, and Department of Physics and Astronomy, Iowa State University, Ames, Iowa 50011, USA}

\date{\today}

\begin{abstract} 

     We report $^{153}$Eu, $^{75}$As and $^{59}$Co nuclear magnetic resonance (NMR) results on  EuCo$_2$As$_2$ single crystal.
     Observations of  $^{153}$Eu and $^{75}$As NMR spectra in zero magnetic field at 4.3 K below an antiferromagnetic (AFM) ordering temperature $T_{\rm N}$ = 45 K and its external magnetic field dependence clearly evidence an incommensurate helical AFM structure  in  EuCo$_2$As$_2$.
     Furthermore, based on $^{59}$Co NMR data in both the paramagnetic and the incommensurate AFM states, we have determined the model-independent value of the AFM propagation vector {\bf k} = (0, 0, 0.73 $\pm$ 0.07)2$\pi$/$c$ where $c$ is the $c$ lattice parameter.  
  Thus the incommensurate helical AFM state was characterized by only NMR data with model-independent analyses, showing NMR to be a unique tool for determination of  the spin structure in incommensurate helical AFMs.

\end{abstract}

 \pacs{75.25.-j, 75.50.Ee, 76.60.-k}

\maketitle

  \section{I. Introduction} 
 
    Understanding the magnetism in $A$Fe$_2$As$_2$ ($A$ = Ca, Ba, Sr, Eu) known as $``$122$"$ compounds with a ThCr$_2$Si$_2$-type structure  at room temperature became one of the important issues after the discovery of iron-pnictide superconductors \cite{Kamihara2008, Johnston2010, Canfield2010, Stewart2011}.
    These systems undergo coupled structural and magnetic phase transitions at a system-dependent N\'eel temperature $T_{\rm N}$, below which long-range stripe-type antiferromagnetic (AFM) order emerges originating from Fe $3d$ electron spins.  
      Superconductivity (SC) in these compounds emerges upon suppression of the stripe-type AFM phase by application of pressure and/or carrier doping.
     Because of the proximity between the AFM and the SC phases, it is believed that stripe-type AFM spin fluctuations play an important role in driving the SC in the iron-based superconductors, although orbital fluctuations are also pointed out  to be important \cite{Kim2013}.
   Recently ferromagnetic (FM) correlations were revealed to also play an important role in the iron-based superconductors \cite{Johnston2010,Nakai2008,PaulPRB,PaulPRL,JeanPRB}.

   EuFe$_2$As$_2$, which exhibits SC under the application of 2--3 GPa of pressure and/or carrier doping \cite{Tereshima2009,Jin2016},  is a special member in the ``122'' class of compounds, as Eu$^{2+}$ has a large magnetic moment ($J = S = $ 7/2, $L$ = 0), where $J$, $S$, and $L$ are the total, spin and orbital angular momenta, respectively.
      EuFe$_2$As$_2$ exhibits the stripe-type AFM order at 186 K due to the Fe spins, while the Eu$^{2+}$ moments order antiferromagnetically below 19 K  with  an A-type AFM structure where the Eu ordered moments are FM aligned in the $ab$ plane but the moments in adjacent layers along the $c$ axis are antiferrmagnetically aligned \cite{Jeevan2008}.
     With substitution of Co atoms for the Fe atoms in Eu(Fe$_{1-x}$Co$_x$)$_2$As$_2$, the ground-state magnetic structure of the Eu$^{2+}$ spins is found  to develop from the A-type AFM order in the parent compound, via the A-type canted AFM structure with some net FM moment component along the crystallographic $c$  direction at intermediate Co doping levels around $x \sim$ 0.1, and then  to the pure FM order along the $c$ axis at $x$ $\sim$ 0.18 (Ref. \onlinecite{Jin2016}).
     With further substitution up to $x$ = 1, EuCo$_2$As$_2$  is reported to again exhibit  A-type AFM order of the Eu ordered moments below $T_{\rm N}$ $\sim$ 40~K \cite{Raffius1993,Ballinger2012}, similar to the parent compound. 
    On the other hand, recent neutron diffraction (ND) measurement reported a planar helical AFM  structure below 47 K where the Eu ordered moments are aligned in the $ab$ plane with the helical axis along the $c$ axis \cite{Tan2016}.
     Therefore, it is important to elucidate the magnetic state of Eu in EuCo$_2$As$_2$ by using different experimental techniques.

    Nuclear magnetic resonance (NMR) is  a powerful technique to investigate magnetic properties of materials from a microscopic point of view. 
   In particular, one can obtain direct and local information of magnetic state at nuclear sites. 
    Although Eu, Co and As are NMR active nuclei in EuCo$_2$As$_2$,  there have been no NMR studies of this compound up to now to our knowledge.

\begin{figure}[tb]
\includegraphics[width=8.5cm]{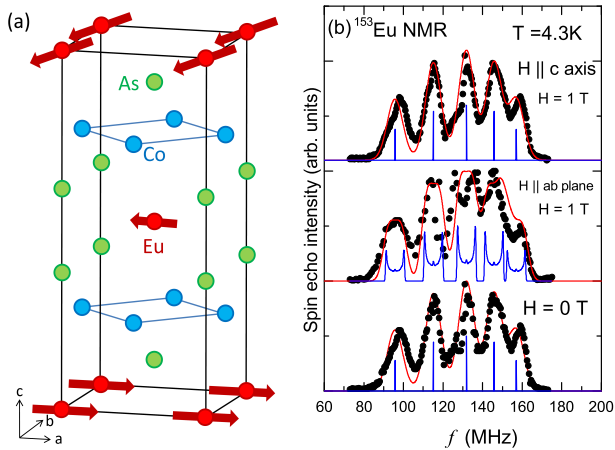} 
\caption{(Color online) (a) Crystal and magnetic structures of EuCo$_2$As$_2$.  (b) $^{153}$Eu-NMR spectra at $T$ = 4.3 K in the AFM state for EuCo$_2$As$_2$ in $H$ = 0  (bottom), $H$ = 1~T parallel to the $ab$ plane (middle) and parallel to the $c$ axis (top). The red and blue lines are the calculated $^{153}$Eu NMR spectra with and without a distribution of $B_{\rm int}^{\rm Eu}$, respectively.
 }
\label{fig:EuNMR}
\end{figure}

    In this paper, we have carried out NMR measurements  to investigate the magnetic and electronic states of each ion in EuCo$_2$As$_2$, where we succeeded in observing NMR signals from all three $^{151}$Eu, $^{59}$Co and $^{75}$As nuclei. 
     From the external field dependence of $^{153}$Eu and $^{75}$As NMR spectra at 4.3 K, below $T_{\rm N}$ = 45 K an incommensurate helical AFM state shown in  Fig.\ \ref{fig:EuNMR}(a)  was clearly evidenced in  EuCo$_2$As$_2$.
    Furthermore, the AFM propagation vector characterizing the helical AFM state is determined to be {\bf k} = (0, 0, 0.73 $\pm$ 0.07)2$\pi$/$c$  from the internal magnetic induction at the Co site obtained by $^{59}$Co NMR under zero magnetic field.
    $^{59}$Co NMR revealed that no magnetic ordering of the Co $3d$ electron spins occurs in the helical AFM state, evidencing that the magnetism in EuCo$_2$As$_2$ originates from only the Eu spins. 
        These results are consistent with the recent neutron diffraction measurements~\cite{Tan2016}.

 \section{II. Experiment}

   A single crystal  ($9\times8\times1$ mm$^3$) of EuCo$_2$As$_2$ for the NMR measurements was grown using Sn  flux \cite{Sangeetha2017}. 
   NMR measurements of $^{153}$Eu ($I$ = $\frac{5}{2}$, $\frac{\gamma_{\rm N}}{2\pi}$ = 4.632 MHz/T, $Q=$ 2.49 barns),   $^{59}$Co ($I$ = $\frac{7}{2}$,  $\frac{\gamma_{\rm N}}{2\pi}$ = 10.03 MHz/T, $Q=$ 0.4 barns), and $^{75}$As ($I$ = $\frac{3}{2}$, $\frac{\gamma_{\rm N}}{2\pi}$ = 7.2919 MHz/T, $Q=$ 0.29 barns) nuclei were conducted using a homemade phase-coherent spin-echo pulse spectrometer. 
   In the AFM state, $^{153}$Eu, $^{75}$As and $^{59}$Co NMR spectra in zero and nonzero magnetic fields $H$ were measured in steps of frequency $f$ by measuring the intensity of the Hahn spin echo. 
   In the paramagnetic (PM) state,  $^{59}$Co NMR spectra were obtained by sweeping the magnetic field at $f$ = 51.1 MHz.
   The $^{75}$As nuclear spin-lattice relaxation rate 1/$T_{\rm 1}$ was measured with a saturation recovery method \cite{T1}.

  \section{III. Results and discussion}
   
      At the bottom panel of Fig.\ \ref{fig:EuNMR}(b) the $^{153}$Eu NMR spectrum in the AFM state for EuCo$_2$As$_2$ is shown, measured in zero magnetic field at a temperature $T$ = 4.3 K.
        The observed spectrum is well reproduced by the following nuclear spin Hamiltonian which produces a spectrum with a central transition line flanked by two satellite peaks on both sides for $I$ = 5/2,  
   ${\cal H} = {\cal -}\gamma\hbar{\bf I}\cdot{\bf B_{\rm int}}+ \frac{h \nu_{\rm Q}}{6} [3I_{z}^{2}-I(I+1) + \frac{1}{2}\eta(I_+^2 +I_-^2)]$, where $B_{\rm int}$ is the internal magnetic induction at the Eu site, $h$ is Planck's constant, and $\nu_{\rm Q}$ is nuclear quadrupole frequency defined by $\nu_{\rm Q} = 3e^2QV_{ZZ}/2I(2I-1)h$  $(=3e^2QV_{ZZ}/20h$ for $I$ = 5/2)  where $Q$ is the electric quadrupole moment of the Eu nucleus, $V_{ZZ}$ is the electric field gradient (EFG) at the Eu site, and $\eta$ is the asymmetry parameter of the EFG \cite{Slichter_book}.
    Since the Eu site in EuCo$_2$As$_2$ has a tetragonal local symmetry (4$/mmm$), $\eta$ is zero. 
   The blue lines shown at the bottom panel of Fig.\ \ref{fig:EuNMR}(b) are the calculated positions for $^{153}$Eu zero-field NMR (ZFNMR) lines using the parameters $|B_{\rm int}^{\rm Eu}|$  = 27.5(1) T, $\nu_{\rm Q}$ = 30.6(1)  MHz and $\theta = 90^\circ$. 
   Here $\theta$ represents the angle between $B_{\rm int}^{\rm Eu}$  and the principle axis of the EFG tensor at the Eu sites. 
   As shown by  the red curve in the figure, the observed $^{153}$Eu ZFNMR spectrum was well reproduced with a broadening of $\sim$ 1.1 T of the calculated lines originated from a distribution of $B_{\rm int}$ probably due to Eu ordered moment distributions.

   Since $B_{\rm int}^{\rm Eu}$ is perpendicular  to the $c$ axis as will be shown below, the principle axis of the EFG  is found to be the $c$ axis, which is similar to the case of the Eu nucleus in EuGa$_4$ with the same ThCr$_2$Si$_2$-type  crystal structure in which the similar values of $B_{\rm int}^{\rm Eu}$ = 27.08~T and  $\nu_{\rm Q}$ = 30.5~MHz for $^{153}$Eu have been reported \cite{Yogi2013}.
         $B_{\rm int}^{\rm Eu}$ is proportional to $A_{\rm hf}$$<$$\mu$$>$ where $A_{\rm hf}$ is the  hyperfine coupling constant and $<$$\mu$$>$ is the ordered Eu magnetic moment. 
      The hyperfine field at the Eu sites mainly originates from core polarization from 4$f$ electrons and is oriented in a direction opposite to that of the Eu moment \cite{Freeman1965}. 
      For $|$$B_{\rm int}^{\rm Eu}$$|$ = 27.5(1)~T and the reported AFM ordered moment $<$$\mu$$>$  = 7.26(8)~$\mu_{\rm B}$/Eu\@ from ND \cite{Tan2016},  $A_{\rm hf}$ is estimated to be  $-$3.78~T/$\mu_{\rm B}$ where the sign is reasonably assumed to be negative due to the core-polarization mechanism. 
      The estimated $A_{\rm hf}$ is not far from  the core-polarization hyperfine coupling constant $-$4.5~T/$\mu_{\rm B}$ estimated for Eu$^{2+}$ ions \cite{Freeman1965}. 
      The small difference could be explained by a positive hyperfine coupling contribution due to conduction electrons which cancel part of the negative core polarization field.  

     In order to determine the direction of $B_{\rm int}^{\rm Eu}$ with respect to the crystal axes, we measured $^{153}$Eu NMR in the single crystal in nonzero $H$. 
   When $H$ is applied along the $c$ axis, almost no change of the $^{153}$Eu NMR spectrum is observed [see the top panel in Fig.~\ref{fig:EuNMR}(b) where the simulated spectra shown by blue and red lines are the same as the case of $H$ = 0]. 
    This indicates that $H$ is perpendicular to the ordered Eu moments and thus to $B_{\rm int}^{\rm Eu}$. 
    Since the effective field at the Eu site is given by the vector sum of  $\bf{B}_{\rm int}^{\rm Eu}$ and $\bf{H}$, i.e., $|$$\bf{B}_{\rm eff}$$|$ = $|$$\bf{B}_{\rm int}^{\rm Eu}$ + $\bf{H}$$|$,   the resonance frequency is expressed for $H$ $\perp$ $<$$\mu$$>$  as $f$ =  $\frac{\gamma_{\rm N}}{2\pi}$$\sqrt {(B_{\rm int}^{\rm Eu})^2+H^2}$. 
    For our applied field range where $B_{\rm int}^{\rm Eu}$ $>>$ $H$, any shift in the resonance frequency due to $H$ would be small, as observed. 
 
    In the case of ${\bf H}$ applied perpendicular to the $c$ axis, on the other hand, each line broadens  as shown in the middle panel of Fig.~\ref{fig:EuNMR}(b).  
   The broadening of each line cannot be explained by the A-type AFM state. 
    In this case,  one expects a splitting of each line into two lines corresponding to two Eu planes where the Eu ordered moments are parallel or antiparallel to ${\bf H}$. 
    In order to explain the observed spectrum, we consider a planar helical structure which produces a two dimensional powder pattern. 
    The blue solid line is a calculated spectrum for an incommensurate helical AFM state. 
     With the inhomogeneous magnetic broadening due to the same distribution of $B_{\rm int}^{\rm Eu}$ as in the $H$ = 0 T spectrum, the observed spectrum at $H$ = 1~T is reasonably reproduced as shown by the red solid curve.
     Thus  these NMR results reveal  an incommensurate helical spin structure with the ordered moments aligned in the $ab$ plane, consistent with recent ND measurements \cite{Tan2016}. 
    The observed $ab$-plane alignment of the ordered moments is also consistent with the prediction of the moment alignment from magnetic dipole interactions between the Eu spins \cite{Johnston2016}.
     A similar incommensurate helical spin structure has been reported in EuCo$_2$P$_2$ \cite{Reehuis1992,Sangeetha2016}.

\begin{figure}[tb]
\includegraphics[width=7.0cm]{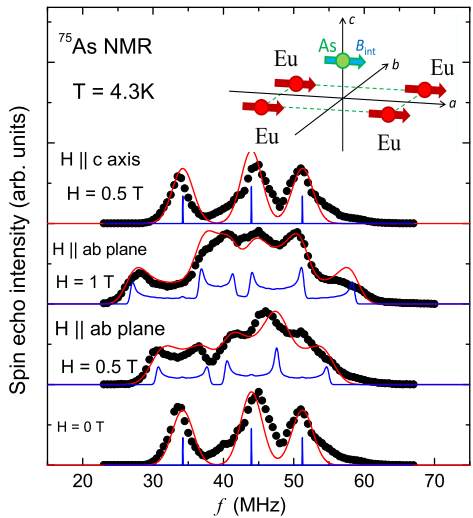} 
\caption{(Color online)  $^{75}$As-NMR spectra at $T$ = 4.3 K in the AFM state for EuCo$_2$As$_2$ in zero magnetic field, and magnetic fields parallel to the $ab$ plane and parallel to the $c$ axis.   
   The red and blue lines are the calculated $^{75}$As-NMR spectra with and without a distribution of $B_{\rm int}^{\rm As}$, respectively. 
  The inset shows the coordination of nearest-neighbor Eu sites around an As site. The arrows on the Eu and As atoms indicate the directions of the Eu ordered  moments and the internal magnetic induction at the As site, respectively.}
\label{fig:AsNMR}
\end{figure}


  The incommensurate planar helical structure is also revealed by $^{75}$As NMR measurements.
    The bottom panel in Fig.~\ \ref{fig:AsNMR} shows the $^{75}$As ZFNMR spectrum at 4.3~K in the AFM state, where the blue lines are the expected positions for the three lines (for $I$ = 3/2) calculated with the parameter $|B_{\rm int}^{\rm As}|$  = 5.86~T, $\nu_{\rm Q}$ = 17.0~MHz and $\theta = 90^\circ $.
    As in the case of the $^{153}$Eu ZFNMR spectrum, the observed $^{75}$As ZFNMR spectrum is well reproduced with an inhomogeneous magnetic broadening of 4~kOe, as shown by the red curve.
     The distribution of $B_{\rm int}^{\rm As}$ originates from the distributions of the Eu ordered moments and its directions. 
         When  ${\bf H}$ is applied along the $c$ axis, almost no change of the spectrum is observed as typically shown in the top panel of Fig.~\ref{fig:AsNMR} where $H$ = 0.5~T.
   This indicates that $\bf H$ is perpendicular to $\bf  B_{\rm int}$ at the As site. 
    On the other hand, when $\bf H$ is applied parallel to the $ab$ plane, similar to the case of $^{153}$Eu ZFNMR spectrum,  each line broadens and exhibits a characteristic shape, again expected for the incommensurate planer helical AFM state.

    According to Yogi {\it et~al}., the direction of $B_{\rm int}^{\rm As}$  is parallel to the Eu ordered moments in the case where the Eu ordered moments are ferromagnetically aligned in the Eu plane \cite{Yogi2013}. 
    Therefore, one can expect almost no change of the $^{75}$As ZFNMR spectrum when $H$ is perpendicular to the Eu ordered moment, as observed in the $^{75}$As ZFNMR spectrum for $H$ $\parallel$ $c$ axis.
    On the other hand,  if one applies $H$ $\parallel$ $ab$ plane, a splitting of the $^{75}$As ZFNMR spectrum is expected similar to the case of the  $^{153}$Eu ZFNMR spectrum. 
    The blue lines in the two middle panels of Fig.~\ref{fig:AsNMR} are calculated spectra of $^{75}$As NMR for the planar helical AFM structure under $H$ = 0.5~T and 1~T. 
    With the same inhomogeneous magnetic broadening ($\sim$ 4~kOe)  due to a distribution of $B_{\rm int}^{\rm As}$, both spectra are well reproduced as shown by the red curves.

\begin{figure}[tb]
\includegraphics[width=8.0cm]{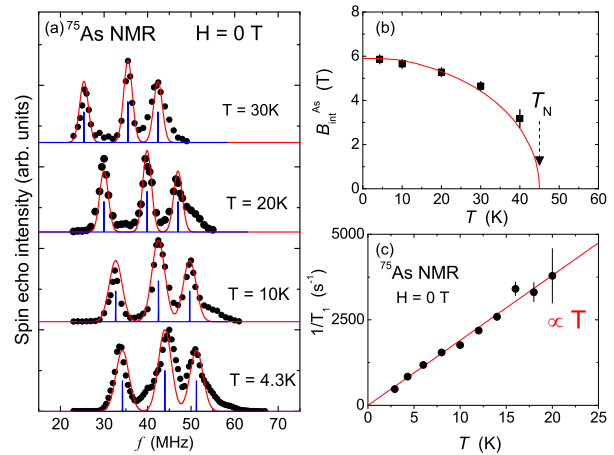} 
\caption{(Color online) (a) Temperature dependence of $^{75}$As-NMR spectra in zero magnetic field in the AFM state. 
  The  red and blue lines  are the calculated $^{75}$As-NMR spectra. 
     (b) Temperature dependence of $B_{\rm int}^{\rm As}$. The solid curve is the Brillouin function with $J$ = $S$ = 7/2.
     (c) Temperature dependence of $^{75}$As-NMR 1/$T_1$ under zero magnetic field. The straight line shows the Korringa relation $1/T_1$ = 190 $T$~(s$^{-1}$).
 }
\label{fig:AsNMR2}
\end{figure}

 The $T$ dependence of the $^{75}$As ZFNMR spectrum is shown in  Fig.~\ref{fig:AsNMR2}(a). 
   With increasing $T$, the spectra shift to lower frequency due to reduction of the internal magnetic induction  $|B_{\rm int}^{\rm As}|$ which decreases from 5.86~T at 4.3~K to 3~T at 40~K. 
     No obvious change in $\nu_{\rm Q}$ = 17.0~MHz  is observed. 
    The $T$ dependence of $|B_{\rm int}^{\rm As}|$  shown in Fig.~\ref{fig:AsNMR2}(b), which is the $T$ dependence of  the order parameter of the planar helical AFM state, is well reproduced by a Brillouin function which was calculated based on the Weiss molecular field model with $J$ = $S$ = 7/2, $T_{\rm N}$ = 45 K and $H_{\rm int}^{\rm As}$ = 5.86 T. 
   This indicates that  the magnetic state of the Eu ions is well explained by the local moment picture although the system is metallic as determined from electrical resistivity measurements \cite{Sangeetha2017}.
The metallic ground state was confirmed by the $T$ dependence of $1/T_1$ measured at the central line of the $^{75}$As-ZFNMR spectrum.
    As shown in  Fig. \ref{fig:AsNMR2}(c),  $1/T_1$ is proportional to $T$,  thus obeying a Korringa law $1/T_1T$ = 190 (sK)$^{-1}$. 
    This confirms a metallic state from a microscopic point of view.


\begin{figure}[tb]
\includegraphics[width=8.0cm]{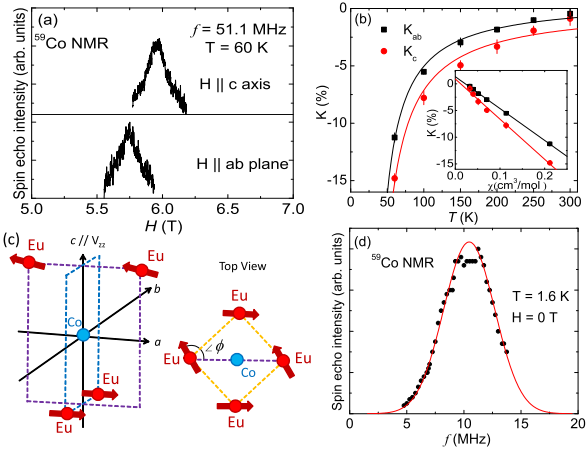} 
\caption{(Color online) (a) Field-swept $^{59}$Co-NMR spectra of EuCo$_2$As$_2$ for $H \parallel ab$ plane (bottom) and $H \parallel c$ axis (top) measured at $f$ = 51.1 MHz and $T$ = 60 K.  
     (b) Temperature dependence of $^{59}$Co Knight shifts  $K_{\rm ab}$ and $K_{\rm c}$ in the paramagnetic state. 
     The solid curves are fits to the data by Curie-Weiss law. 
      The inset shows $K(T)$ versus magnetic susceptibility $\chi(T)$ for the corresponding $ab$ and $c$ components of $K$.  The solid  lines are linear fits.
    (c) Coordination of nearest-neighbor Eu sites around Co site. The arrows on the Eu atoms indicate the ordered magnetic moments. The magnetic moment turn angle between adjacent magnetic layers is $\phi$.  
    (d) $^{59}$Co-NMR spectrum at $T$ = 1.6 K in the AFM state in zero magnetic field. 
 }
\label{fig:CoNMR}
\end{figure}

  Now we discuss our $^{59}$Co NMR data for both the PM and AFM ordered states.
    Figure \ref{fig:CoNMR}(a) shows the  field-swept $^{59}$Co NMR spectra in the PM state at $T$ = 60~K for $H \parallel c$ and $H \parallel ab$.
    For $I$ = 7/2 nuclei, one expects a central transition line with three satellite lines on both sides.   
    The observed spectra, however, do not show the seven distinct lines but rather exhibit a single broad line due to inhomogeneous magnetic broadening.
     The $T$ dependence of the NMR shift for $H$ $\parallel$ $c$ ($K_{c}$) and $H$ $\parallel$ $ab$ ($K_{ab}$) is shown in Fig. \ref{fig:CoNMR}(b), where we fit the data with the Curie-Weiss law $\frac{C}{T-\theta_{\rm p}}$. 
     The solid curves are fits with $C$ = $-$553 ($-$440) $\% \rm K$ and $\theta_{\rm p}$ = 24.0 (24.0) K for $K_{c}$ ($K_{ab}$).  
     The values $\theta_{\rm p}$ = 24.0 K for both field directions indicate predominant FM exchange interactions between Eu spins. 
   This is consistent with the in-plane FM exchange interactions responsible for the planar helical AFM structure.

    The hyperfine coupling constants  $A_{ab}$  and $A_c$ for $^{59}$Co surrounded by four Eu$^{2+}$ ions can be estimated 
from the slopes of $K$-$\chi$ plots with the relation   $A$   = $\frac{N_{\rm A}}{Z}\frac{K(T)}{\chi(T)}$,   
 where  ${N_{\rm A}}$ is Avogadro's number and $Z=4$ is the number of nearest-neighbor Eu$^{2+}$  ions around a Co atom.    
      As shown in the inset of Fig.\ \ref{fig:CoNMR}(b), both  $K_ {ab}$ and $K_c$ vary linearly with the corresponding $\chi$. 
   From the respective slopes, we estimate $A_{ab}$ = ($-$0.875 $\pm$ 0.09) kOe/$\mu_{\rm B}$/Eu  and $A_{c}$ = ($-$1.09 $\pm$ 0.17) kOe/$\mu_{\rm B}$/Eu, respectively.
   These values are much smaller than a typical value $A$  = $-$105  kOe/$\mu_{\rm B}$ for Co $3d$ electron core polarization \cite{Freeman1965}.  
   This indicates that the hyperfine field at the Co site originates from the transferred hyperfine field produced by the Eu$^{2+}$ spins and that no $3d$ spins on the Co sites contribute to the magnetism of EuCo$_2$As$_2$.

  We now consider the influence of the planar helical AFM state on the Co NMR data.
   We have succeeded in observing the $^{59}$Co ZFNMR spectrum  at 4.3 K as shown in Fig. \ref{fig:CoNMR}(d), where the internal magnetic induction at the Co site is estimated to be  $|B_{\rm int}^{\rm Co}|$  = 10.3 kOe. 
     Based on the analysis for $B_{\rm int}$ by Yogi  ${\it et~al.}$ for EuGa$_4$ (Ref. \onlinecite{Yogi2013}), we extended their calculation of  $B_{\rm int}$ to an incommensurate helical AFM state and found  that $|B_{\rm int}|$  at the Co site appears in only the $ab$ plane when the Eu ordered moments lie in the $ab$ plane and is expressed by 
\begin{eqnarray}
 B_{\rm int}^{\rm Co} =  2 \langle \mu \rangle A_{ab}\sqrt{2+2\cos\phi}
\label{eqn:Co_internal}
\end{eqnarray}
where $\phi$ is the turn angle along the the $c$ axis between the Eu ordered moments in adjacent Eu planes, which characterizes the helical structure. 
  In the case of  $\phi$ = $\pi$ corresponding to an A-type collinear AFM state,  $ B_{\rm int}^{\rm Co} $ is zero due to a cancellation of the internal magnetic induction from the four nearest-neighbor Eu ordered moments.  
   On the other hand, if $\phi$ deviates from $\pi$ corresponding to a helical state, one can expect a finite $ B_{\rm int}^{\rm Co}$.  
   Thus the observation of the finite $ B_{\rm int}^{\rm Co}$  is direct evidence of the planar incommensurate helical AFM state in EuCo$_2$As$_2$.
   Furthermore, using Eq.~(\ref{eqn:Co_internal}), we can determine the AFM propagation vector {\bf k} = (0, 0, $k$)2$\pi$/$c$, where $c$ is the $c$-axis lattice parameter of the body-centered tetragonal Eu sublattice.  
    Since the distance $d$ along the $c$ axis between adjacent layers of FM-aligned Eu moments is $d$ = $c$/2,  the turn angle between the ordered moments in adjacent Eu layers is $\phi$ = $kd$, as shown in Fig.\ \ref{fig:CoNMR}(c).
    Using $\langle$$\mu$$\rangle$ = 7.26(8) $\mu_{\rm B}$ \cite{Tan2016},  $A_{ab}$ = $-$0.875  kOe/$\mu_{\rm B}$/Eu and  $B_{\rm int}^{\rm Co}$ = 10.3 kOe, the turn angle $\phi$ is estimated to be 132$^{\circ}$ corresponding to a helical wave vector  {\bf k} = (0, 0, 0.73 $\pm$ 0.07)2$\pi$/$c$.
   This value of {\bf k}  is in very good agreement with ${\bf k}$ = (0, 0, 0.79)2$\pi$/$c$ obtained from ND data \cite{Tan2016}.

   \section{IV. Summary}

     In summary,  we have shown that by analyzing the NMR spectrum in zero field and its external-field dependence, one can determine directly an incommensurate helical AFM structure in EuCo$_2$As$_2$. 
    The AFM propagation vector characterizing the incommensurate helical AFM state was determined model-independently to be {\bf k} = (0, 0, 0.73 $\pm$ 0.07)2$\pi$/$c$ from the internal magnetic field at the Co site obtained by $^{59}$Co NMR under zero magnetic field. 
     Thus  NMR can be a unique tool for a model-independent determination of  the spin structure in incommensurate helical antiferromagnets.
     This should prove valuable for the future investigation of local spin configurations in other europium compounds such as  in EuCo$_2$P$_2$ which is also reported to exhibit an incommensurate helical AFM structure below 66~K \cite{Reehuis1992,Sangeetha2016}. 
   Our NMR approach can also be used to study in detail the magnetism originating from the Eu spins in Eu(Fe$_{1-x}$Co$_x$)$_2$As$_2$ SCs where the magnetic structure of the Eu spins changes from the A-type AFM state to a canted AFM state, and then to the ferromagnetic state with increasing Co substitution \cite{Jin2016}. 
   Such a detailed study would provide clues about the origin of the coexistence of SC and magnetism in the Eu(Fe$_{1-x}$Co$_x$)$_2$As$_2$ system.

   \section{V. Acknowledgments} 

    The authors thank Mamoru Yogi at University of the Ryukyus for helpful discussions.   
    The research was supported by the U.S. Department of Energy, Office of Basic Energy Sciences, Division of Materials Sciences and Engineering. Ames Laboratory is operated for the U.S. Department of Energy by Iowa State University under Contract No.~DE-AC02-07CH11358.
    N. H. thanks the Japan Society for the Promotion of Science KAKENHI : J-physics (Grant Nos. JP5K21732, JP15H05885, and JP16H01078) for financial support to be a visiting scholar at the Ames Laboratory.


\begin{thebibliography}{99}
\bibitem{T1}   Nuclear spin-lattice relaxation rate $1/T_1$ at each $T$ was determined by fitting the nuclear magnetization $M$ versus time $t$ using the exponential function $1-M(t)/M(\infty) = 0.1 e^ {-t/T_{1}} +0.9e^ {-6t/T_{1}}$ for $^{75}$As NMR, where $M(t)$ and $M(\infty)$ are the nuclear magnetization at time $t$ after saturation and the equilibrium nuclear magnetization at $t$ $\rightarrow$ $\infty$, respectively.
\end{thebibliography}
\end{document}